\documentclass{book}

\usepackage[paperwidth=17cm, paperheight=24cm, includeheadfoot,
            left=2cm, right=2cm, top=2cm, bottom=1.3cm, twoside]{geometry}

\usepackage
 {amsmath,amssymb,caption,fancyhdr,graphicx,lineno,remreset,sidecap,url}

\def\thepart{\Alph{part}}

\makeatletter
\@removefromreset{figure}{chapter}
\makeatother

\makeatletter
\renewcommand{\thefigure}{\@arabic\c@figure}
\makeatother

\makeatletter
\@removefromreset{table}{chapter}
\makeatother

\makeatletter
\renewcommand{\thetable}{\@arabic\c@table}
\makeatother

\makeatletter
\@removefromreset{equation}{chapter}
\makeatother

\makeatletter
\renewcommand{\theequation}{\@arabic\c@equation}
\makeatother

\usepackage{minitoc}

\newcommand{\Author}{}
\newcommand{\AuthorLastName}[1]{\renewcommand{\Author}{#1}}

\def\TitleAuthor#1#2{\chapter[\thepart\thelecture\ #1
 \textit{(#2)}]{\Large #1}}

\newsavebox{\shortTitleBox}
\def\shortTitleAuthor#1#2{\savebox{\shortTitleBox}{#1 \textit{(#2)}}}

\newcounter{lecture}[part]

\fancypagestyle{plain}
 {
  
  \fancyhead[OL]
   {
    \vspace{-8.5pt}
    \textit{Lecture given at the International Summer School
   \emph{Modern Computational Science}}\\
    \textit{(August 16-28, 2009, Oldenburg, Germany)}
  }
 }

 {
  
  \fancyhf{}
  \fancyhead[OR]{\rightmark}
  \fancyfoot[OR]{\thepage}
  \fancyfoot[EL]{\thepage}
  \fancyhead[EL]{\usebox{\shortTitleBox}}
 }

\makeatletter
\def\thebibliography#1
{
 \section*{References\@mkboth{References}{References}}
 \list{[\arabic{enumi}]}
  {
   \settowidth\labelwidth{[#1]}\leftmargin\labelwidth\advance\leftmargin
   \labelsep\usecounter{enumi}
  }
 \def\newblock{\hskip .11em plus .33em minus .07em}
 \sloppy\clubpenalty4000\widowpenalty4000\sfcode`\.=1000\relax
}
\makeatother

\newtheorem{Theorem}{Theorem}[section]

\begin{document}

\dominitoc

\faketableofcontents

\captionsetup{width=0.9\textwidth,font=small,labelfont=bf}


\newcommand*{\bigO}[1]{\mathcal{O}\left(#1\right)}
\newcommand*{\I}{\mathrm{i}}
\newcommand*{\e}{\mathrm{e}}
\newcommand*{\F}{\mathbb{F}}
\newcommand{\todo}[1]{\textcolor[rgb]{1,0,0}{\textbf{\emph{#1}}}}
\newcommand{\bigo}{\ensuremath{\mathcal{O}}}

\stepcounter{lecture}
\setcounter{figure}{0}
\setcounter{equation}{0}
\setcounter{table}{0}

\AuthorLastName{Mertens}

\TitleAuthor{Random Number Generators: A Survival Guide for Large Scale Simulations}{Stephan Mertens}

\shortTitleAuthor{Random Number Generators}{Mertens}

{\itshape Stephan Mertens}
\bigskip

\noindent
\parbox[t]{20em}
 {
  \begin{raggedright}
    \itshape
    Institute for Theoretical Physics\\
    Otto-von-Guericke University Magdeburg\\
    D-39106 Magdeburg\\
    Germany
  \end{raggedright}
 }
\parbox[t]{15em}
 {
  \begin{raggedright}
    \itshape 
    Santa Fe Institute\\
    1982 Hyde Park Road\\
    Santa Fe, NM 87501\\
    U.S.A.
  \end{raggedright}
 }

\paragraph{Abstract.} Monte Carlo simulations are an important tool in
statistical physics, complex systems science, and many other fields.
An increasing number of these simulations is run on parallel systems ranging
from multicore desktop computers to supercomputers with thousands of CPUs.
This raises the issue of generating large amounts of random numbers
in a parallel application.
In this lecture we will learn just enough of the theory of pseudo
random number generation to make wise decisions on how to choose
and how to use random number generators when it comes to large scale,
parallel simulations.

\minitoc


\section{Introduction}

The Monte Carlo method is a major industry and random numbers are its key
resource. In contrast to most commodities, quantity and quality of
randomness have an inverse relation: The more randomness you consume,
the better it has to be. The quality issue arises because simulations
only \emph{approximate} randomness by generating a stream of
deterministic numbers, named \emph{pseudo}-random numbers, with
successive calls to a subroutine called pseudo-random number generator (PRNG).
Due to the ever increasing computing power, however, the quality
of PRNGs is a moving target.  Simulations that consume $10^{10}$
pseudo-random numbers were considered large-scale Monte Carlo
simulation a few years ago. On a present-day desktop machine this is a
10-minute run.

More and more large scale simulations are run on parallel systems like
multicore computers, networked workstations or supercomputers with
thousands of CPUs. In a parallel environment the quality of a PRNG is
even more important, to some extent because feasible sample sizes are
easily $10\ldots 10^5$ times larger than on a sequential machine.  The
main problem is the parallelization of the PRNG itself, however.  Some
good generators are hardly parallelizable, others lose their
efficiency, their quality or even both when being parallelized.

In this lecture we will discuss the generation of random numbers with
a particular focus on parallel simulations. We will see that there are
ready-to-use software packages of parallelizable PRNGs, but the proper
use of these packages requires some background knowlegde. And this
knowledge is provided in this lecture.

\section{Recurrent Randomness}
\label{sec:basics}

\begin{figure}
  \centering
  \includegraphics[width=0.3\columnwidth]{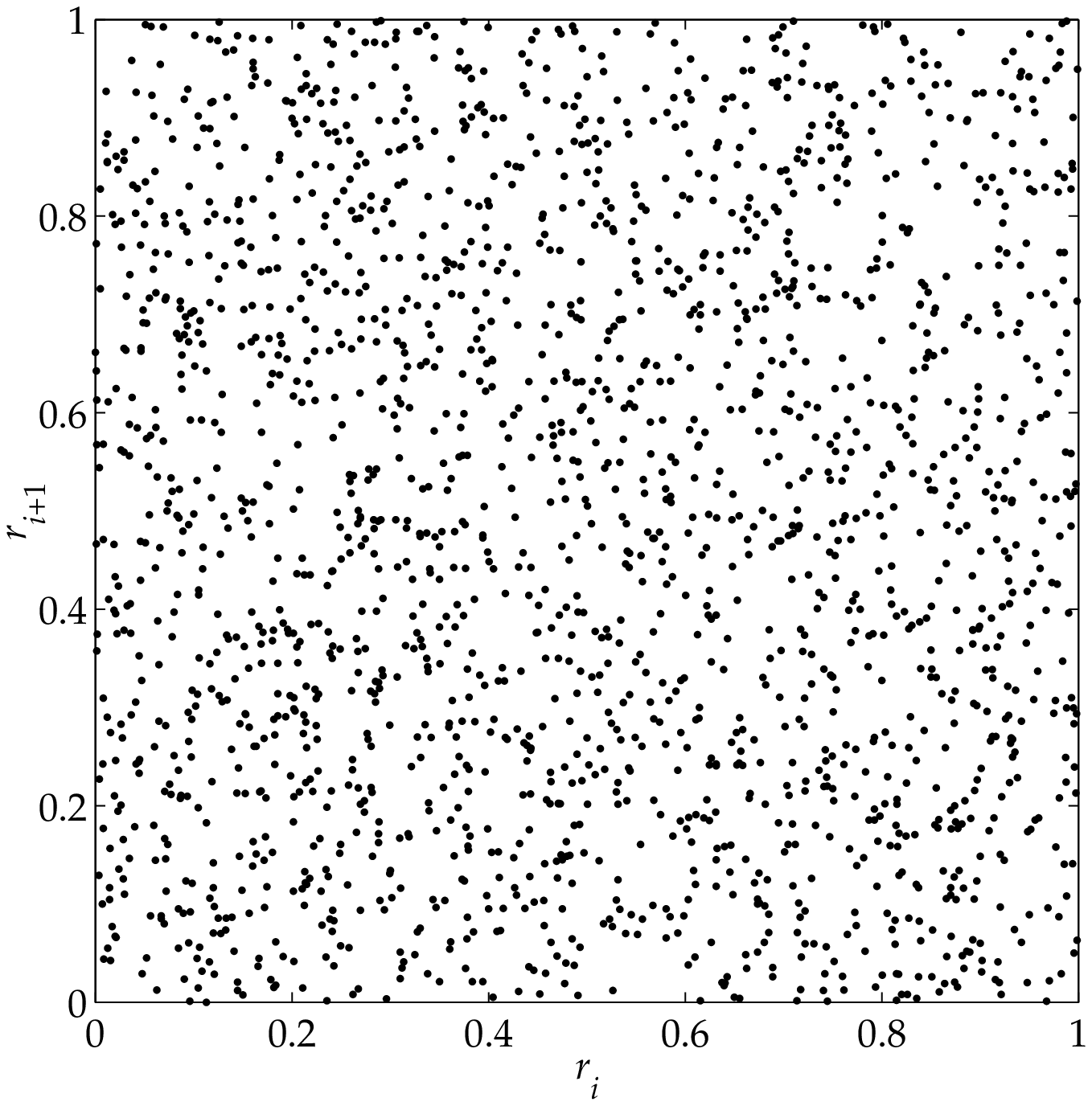}
  \includegraphics[width=0.3\columnwidth]{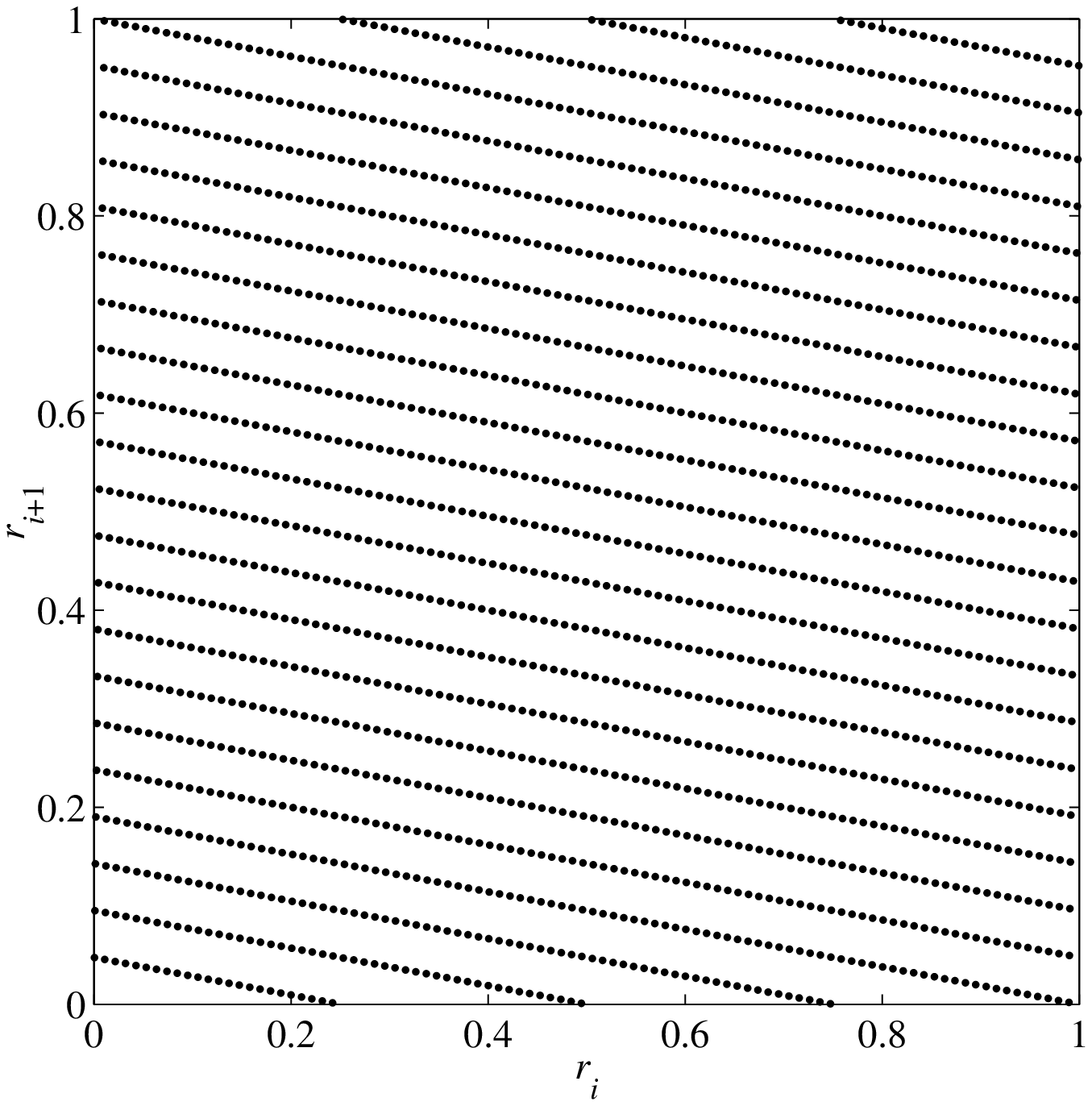}
  \includegraphics[width=0.3\columnwidth]{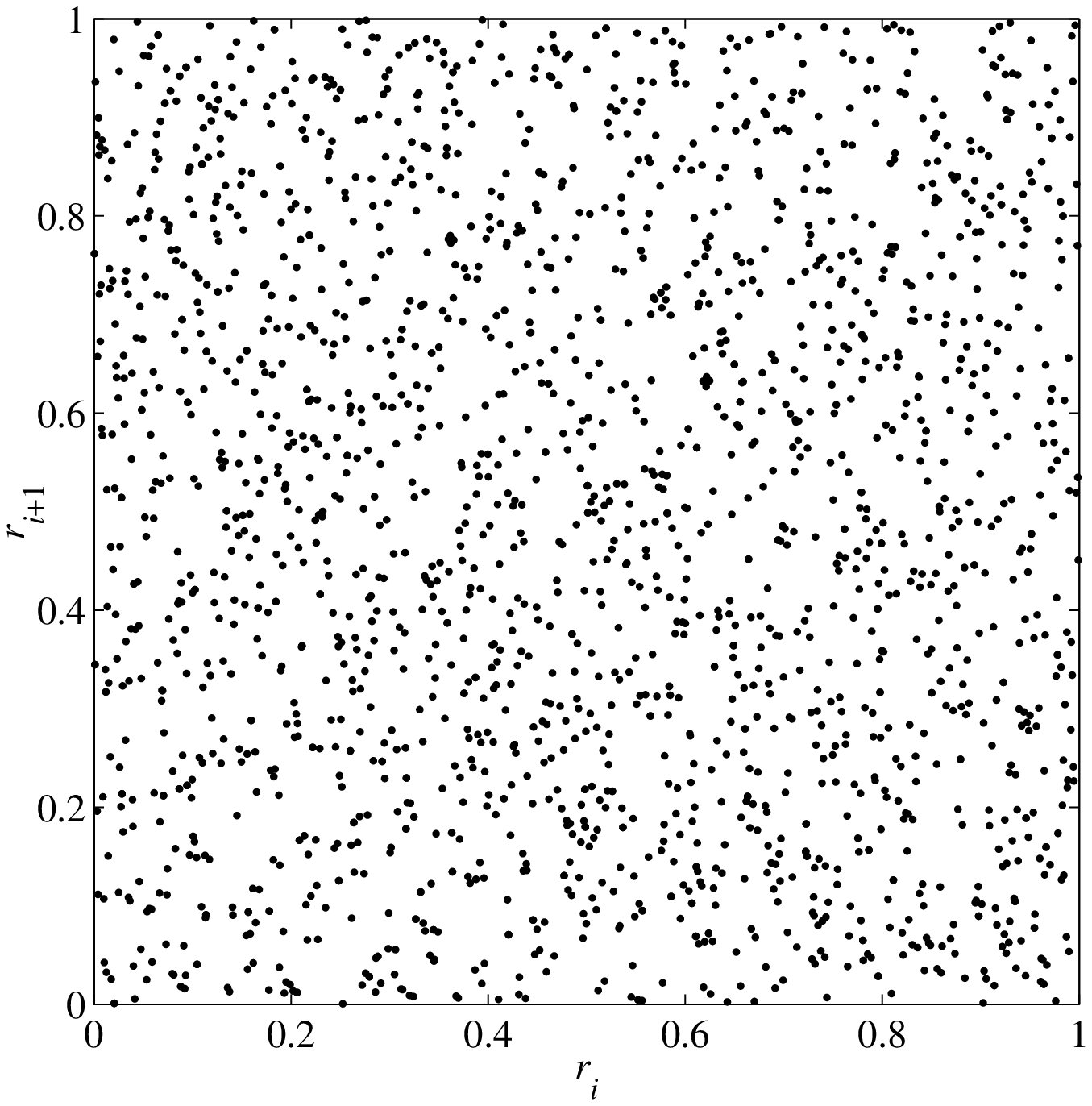}
  \caption{Points in the unit square, generated from different (pseudo)random sources.
  \label{fig:examples}}
\end{figure}

Most programming languages come with a command like \texttt{rand()}
that returns a ``random'' number each time it is invoked. How does
this work? One possible mechanism is that \texttt{rand()} accesses
some device in the computer that produces random bits, pretty much
like \texttt{time()} accesses the hardware clock to read the current
time. In fact there are electronic devices designed to produce
randomness. In these devices, the thermal noise in a resistor or the
shot noise of a diode is measured and turned into a stream of random
bits. On Unix machines, there is a device called \texttt{/dev/random}
that returns ``random'' numbers based on environmental noise like
keystrokes, movements of the mouse, or network traffic.
The leftmost set of points in Figure~\ref{fig:examples} has been generated
with random numbers from \texttt{/dev/random}.

These more or less fundamental, physical sources of randomness are not
behind functions like \texttt{rand()}, however. The randomness used in
simulations is generated \emph{mathematically}, not physically. And this is
where the term ``pseudo'' comes from. Contrary to the impression
you might have gotten in high school, mathematical procedures
are deterministic, not random. So why do people not use
physical noise in simulations?

One reason is speed: accessing an external device and postprocessing
the noisy signal is much slower than having the CPU perform a few
simple calculations. Another reason is reproducibility. With physical
noise, each run of a simulation yields a different result. But a
numerical experiment like a simulation should be reproducible. You
might object that real, physical experiments are always subject to
uncontrollable noise. That's right, but the experimentalists work hard
to reduce this noise as much as possible. And the advantage of
numerical experiments is that we can control everything, including the
noise. And we don't want to give up this position. And, of course,
reproducibility is mandatory when it comes to debugging.

So how does a purely mathematical random number generator work?
The main idea is to generate 
a sequence $(r) = r_1,r_2,\ldots$ of pseudo-random
numbers by a recurrence 
\begin{equation}
  \label{eq:recurrence}
  r_i = f(r_{i-1}, r_{i-2}, \ldots, r_{i-n})\,,
\end{equation}
and the art of random number generation lies in the design of the
function $f$. Note that we need to provide the first $n$ numbers to
get this recurrence off the ground. This is called \emph{seeding}, and
your favourite programming language has a command for this, like
\texttt{srand(seed)} in C. Since \eqref{eq:recurrence} is
deterministic, the only randomness involved is the choice of the seed, which
is then ``spread out'' over a whole, usually very long sequence of
numbers. It is quite surprising that this bold approach actually works!

\subsection{Linear Recurrences and Randomness}

We can mimic true randomness by pseudo randomness well enough,
provided our recurrence is properly designed.  One method to generate
pseudo random integers between $0$ and some prime number $p$ is
the linear recurrence
\begin{equation}
  \label{eq:lfsr}
  r_i = a_1 r_{i-1} + a_2 r_{i-2} + \ldots + a_n r_{i-n} \bmod p\,.
\end{equation}
Here it is the $\bmod$ operation that introduces some randomness
by mimicking the circular arrangement of a roulette wheel.
The quality of this method depends on the magic numbers
$a_1,\ldots,a_k$, and to some extent on $n$ and $p$. Sequences generated by
\eqref{eq:lfsr} are called linear feedback shift register sequences, or
LFSR sequences for short.

\begin{figure}
  \centering
  \includegraphics[width=0.4\linewidth]{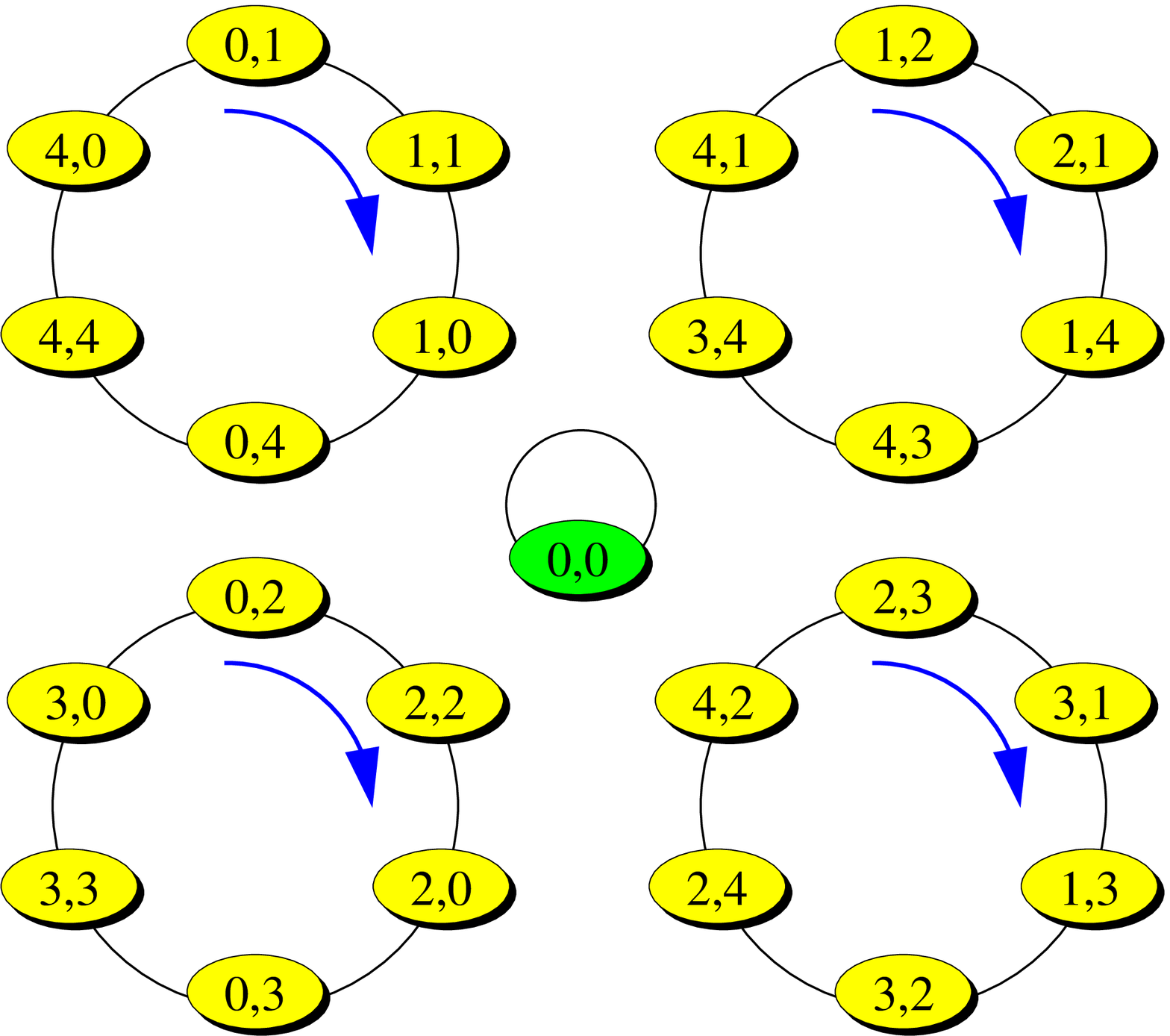}
  \hspace{0.15\linewidth}
  \includegraphics[width=0.4\linewidth]{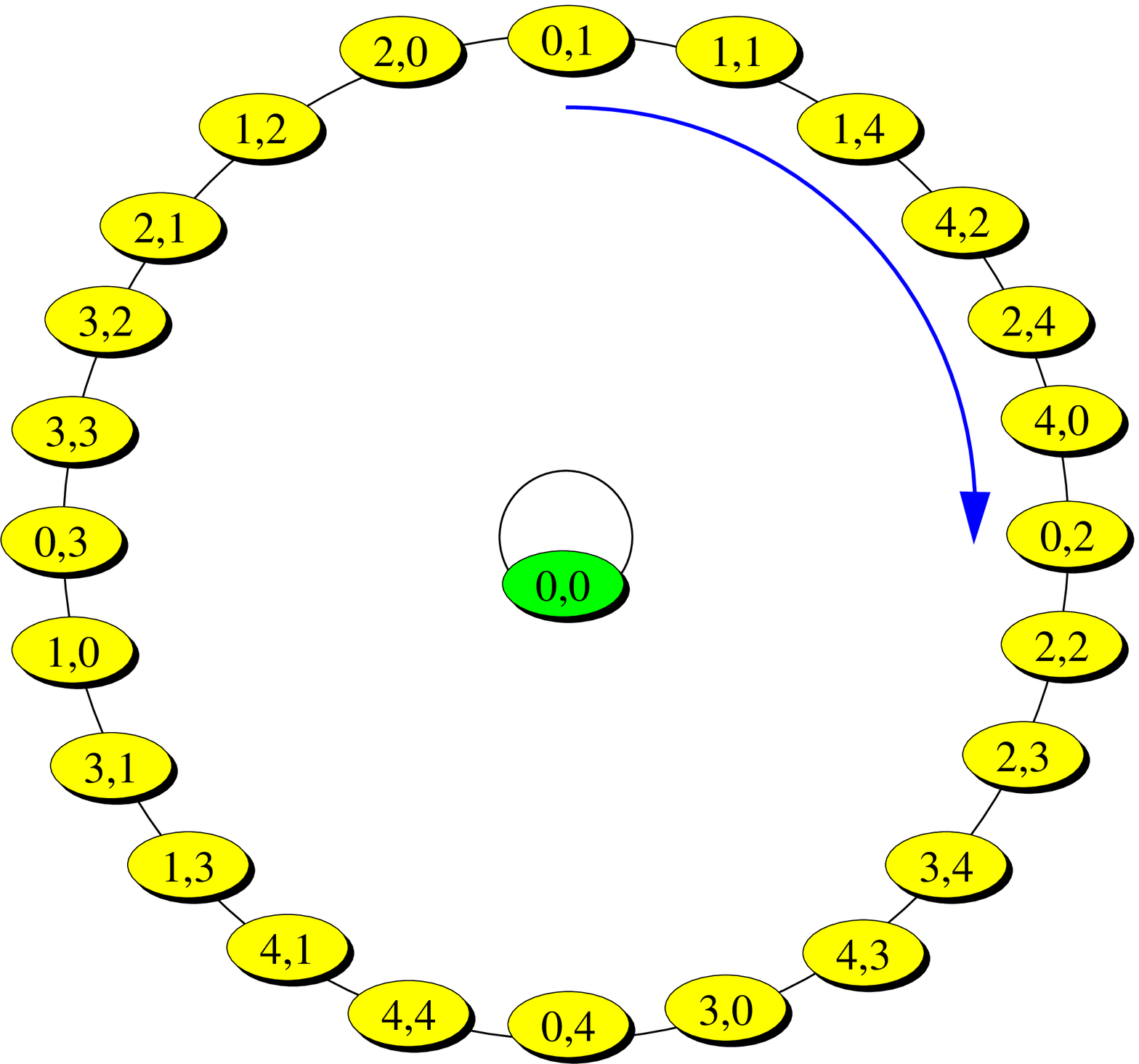}
  \caption{Sequences generated by $x_k = x_{k-1}+4x_{k-2}\bmod 5$ (left) and $x_k=x_{k-1}+3x_{k-2}\bmod 5$ (right).}
  \label{fig:cycles}
\end{figure}

Obviously, any sequence generated by \eqref{eq:lfsr} is periodic.
As a basic requirement we want the period of the sequence to be as
large as possible.
The tuple $(r_{i-1},\ldots,r_{i-n})$ can take on $p^n$ different
values, and the all zero tuple is a fixed point. Hence the maximum period of
the sequence is 
\begin{equation}
  \label{eq:max-T}
  T =p^n-1\,.
\end{equation}
Can we choose the coefficients $a_i$ such that the period takes on the
maximum value?

As a toy example consider the LFSR sequence 
\begin{displaymath}
  x_k = x_{k-2}+4x_{k-1} \bmod 5\,.
\end{displaymath}
The configuration space of this recurrence is composed of $4$ sequences
of period $T=6$ each, and the fixed point $(0,0)$ (Figure~\ref{fig:cycles}).
The slightly modified recurrence
\begin{displaymath}
   x_k = x_{k-2}+3x_{k-1} \bmod 5
\end{displaymath}
achieves the maximum period $T=5^2-1$ and traces the whole
configuration space except $(0,0)$ (Figure~\ref{fig:cycles}).

The theory behind this requires some bits of the
mathematics of finite fields. Remember that
the set of integers $\{0,\ldots,p-1\}$ together with addition and
multiplication modulo $p$ form a finite field $\F_p$ if $p$ is prime.
The corresponding theorem is this:
\begin{Theorem}
  \label{the:period}
  The LFSR sequence \eqref{eq:lfsr} has maximum period
  $p^n-1$ if and only if the characteristic polynomial
  \begin{equation}
    \label{eq:feed-poly}
    f(x) = x^n - a_1 x^{n-1} - a_2 x^{n-2} - \ldots - a_n
  \end{equation}
  is \emph{primitive} modulo $p$.
\end{Theorem}
A monic polynomial $f(x)$ of degree $n$ over $\F_p$ is primitive
modulo $p$, if it is irreducible (i.e., cannot be factorized over
$\F_p$), and if it has a primitive element of the extension field
$\F_{p^n}$ as one of its roots. An element $\alpha$ of a field $\F$ is
primitive if the powers of $\alpha$ generate $\F\setminus 0$.
Primitive polynomials are not hard to find, and there are plenty of
them for given values of $p$ and $n$.

In our toy example above, $f(x) = x^2-x-4$ is irreducible, but not
primitive. Let $\alpha$ be a root of $f$, i.e., $\alpha^2=\alpha+4$.
Then
\begin{center}
\begin{tabular}{lll}
  $\alpha^1 = \alpha$ & $\alpha^2 = 4+\alpha$ & $\alpha^3 = 4$ \\
  $\alpha^4 = 4\alpha$ & $\alpha^5 = 1+4\alpha$ & $\alpha^6 = 1$ 
\end{tabular}
\end{center}
hence $\alpha$ does generate only 6 of the 24 elements of $\F_{5^2}\setminus 0$. Here the order of $\alpha$ equals the period of the LFSR sequence, and this is true in general. 

The roots of the irreducible $f(x) = x^2-x-3$ are primitive (exercise), hence
the corresponding LFSR sequence has maximum period.

The following two theorems tell us that LFSR sequences
with maximum period share some features with
sequences of truly random numbers.
\begin{Theorem}
  \label{the:pn}
  Let $(r)$ be a LFSR sequence \eqref{eq:lfsr} with period
  $T=p^n-1$.  Then each $k$-tuple $(r_{i+1}, \ldots,
  r_{i+k})\in\{0,1,\ldots,p-1\}^k$ occurs $p^{n-k}$ times per period
  for $k \leq n$, except the all zero tuple for $k=n$, which never
  occurs.
\end{Theorem}
If a tuple of $k$ successive terms
$k\leq n$ is drawn from a random position of $(r)$, the
outcome is uniformly distributed over all possible $k$-tuples in
$\F_p$. This is exactly what one would expect from a truly random
sequence.  

\begin{Theorem}
  \label{the:cor}
  Let $(r)$ be a LFSR sequence \eqref{eq:lfsr} with period $T = p^n-1$
  and let $\alpha$ be a complex $p$th root of unity and $\overline{\alpha}$
  its complex conjugate. Then 
  \begin{equation}
    \label{eq:autocorr}
    C(h) := \sum_{i=1}^{T} \alpha^{r_i}\cdot\overline{\alpha}^{r_{i+h}} =
    \begin{cases}
      T & \text{if $h = 0 \bmod T$} \\
     -1 & \text{if $h \neq 0 \bmod T$}
    \end{cases}\,.
  \end{equation}
\end{Theorem}
$C(h)$ can be interpreted as the autocorrelation function of the
sequence, and Theorem~\ref{the:cor} tells us that LFSR sequences
with maximum period have a flat autocorrelation function. Again
this is a property (white noise) that is also observed in truly 
random sequences.

Theorems \ref{the:pn} and \ref{the:cor} are the reason why LFSR
sequences with maximum period are called \emph{pseudonoise
  sequences}. As you can guess by now, they are excellent recipes 
for PRNGs.

\subsection{Yet Another Random Number Generator}

If you check the internet or consult textbooks on random number
generation, you will usually find all sorts of \emph{nonlinear}
generators. The rationale for going nonlinear is that linear
recurrences have some known weaknesses.  Another
motivation is the belief that more complicated rules generate more
random output. If you agree with the latter statement, have a look at
\cite{Knuth:1998:TAoCPII}, in which Donald Knuth tells the story how
he designed a tremendously involved algorithm which one can't even
write down as a simple recurrence. His goal was to create an ``extremely
random'' PRNG, but to his surprise, his rule runs into a fixed point
after just a few iterations. It was completely useless as a PRNG. The
moral of this story:
\begin{quote}
  Random number generators should not be chosen at random.
\end{quote}
Some theory should be used. Unfortunately, the theory of nonlinear
recurrences is much less developed than the theory of linear
recurrences. So let's stick for the moment with linear pseudonoise
sequences and have a look at their weaknesses.

The points in the middle of
Figure~\ref{fig:examples} have been generated by taking two
consecutive and properly scaled values of the linear recurrence
\begin{equation}
  \label{eq:lcg-example}
  r_{i+1} = 95 r_i \bmod 1999
\end{equation}
as coordinates. This is a pseudonoise sequence, but it
doesn't look random at all, at least not in this experiment. The
linear structure that survives the $\bmod$-operation is clearly
visible.  We can get rid of this effect by using a linear recurrence
with more feedback taps, i.e., with $n>1$. But the linear structure
reappears when we sample points in $d$ dimensional
space with $d > n$. 

Theorem \ref{the:pn} tells us that pseudonoise sequences, when used to
sample coordinates in $d$-dimensional space, cover every point for $d
< n$, and every point except $(0,0,\ldots,0)$ for $d=n$. For $d > n$
the set of positions generated is obviously sparse, and the linearity
of the rule (\ref{eq:lfsr}) leads to the concentration of the sampling
points on $n$-dimensional hyperplanes.  This phenomenon, first noticed
by Marsaglia in 1968 \cite{Marsaglia:1968:planes}, motivates one of
the well known empirical tests for PRNGs, the so-called spectral test
\cite{Knuth:1998:TAoCPII}. The spectral test checks the behavior of a
generator when its outputs are used to form $d$-tuples.  Linear
sequences can fail the spectral test in dimensions larger than the
register size $n$.  Closely related to this mechanism are the observed
correlations in other empirical tests like the birthday spacings test
and the collision test.  Nonlinear generators do quite well in all
these tests, but compared to linear sequences they have much less nice
and \emph{provable} properties. And they cannot be properly parallelized,
as we will see below.

To get the best of both worlds we can use a delinearization that preserves
all the nice properties of linear pseudonoise sequences:
\begin{Theorem}
  \label{the:yarn}
  Let $(q)$ be a linear pseudonoise sequence in $\F_p$, and let $g$ be
  a generating element of the multiplicative group $\F^*_p$. Then the
  sequence $(r)$ with
  \begin{equation}
    \label{eq:def-yarn}
    r_i = 
    \begin{cases}
      g^{q_i}\bmod p & \text{if $q_i > 0$} \\
      0 & \text{if $q_i = 0$}
    \end{cases}
  \end{equation}
  is a pseudonoise sequence, too.
\end{Theorem}
The proof of this theorem is trivial: since $g$ is a generator of $\F^*_p$,
the map \eqref{eq:def-yarn} is bijective. We call a generator based on \eqref{eq:def-yarn} YARN generator (yet another random number).

The exponentiation largely destroys the linear structure. This can be seen
in the rightmost plot in Figure~\ref{fig:examples}, which has been generated
by
\begin{displaymath}
  \begin{aligned}
    q_{i} &= 95 q_{i-1} \bmod 1999 \\
    r_{i} &= 1099^{q_i} \bmod 1999\,.
  \end{aligned}   
\end{displaymath}
Visually, this delinearized sequence scatters the points much more
randomly than the underlying linear version. In fact, YARN type
generators pass the corresponding empirical tests (spectral tests,
coupon collector, etc.)  with flying colors, but there is an even more
convincing argument that exponentiation removes all traces of
linearity. This argument is based on the following theorem:
\begin{Theorem}
  \label{the:linear_recurrences}
  All finite length sequences and all periodic sequences over a finite field
  $\F_p$ can be generated by a linear recurrence \eqref{eq:lfsr}.
\end{Theorem}
This theorem implies that \emph{any} recurrent sequence of integer
numbers of a finite domain, i.e., every sequence that a computer
program can generate, is equivalent to some linear recurrence in
$\F_p$. In this sense, pseudorandomness is always linear.

Prima facie, Theorem \ref{the:linear_recurrences} is counterintuitive.
But note that it says nothing about order of the linear recurrence.
The \emph{linear complexity} of a sequence is the minimum order of a
linear recurrence that generates this sequence. Consider the hardest
case: a sequence of $\ell$ truly random numbers between $0$ and $p-1$.
A truly random sequence of $\ell$ numbers cannot be compressed, i.e., 
expressed by less than $\ell$ numbers. Hence we expect the linear complexity
of $\ell$ random numbers to be $\ell/2$: half of the information is contained in 
the coefficients, the other half in in the seed. In fact one
can prove that the average linear complexity of an $\ell$ element random
sequence is $\ell/2$. 

Typically the linear complexity of a nonlinear sequence is of the same
order of magnitude as the period of the sequence.  Since the designers
of PRNGs strive for ``astronomically'' large periods, implementing
nonlinear generators as a linear recurrences is not tractable.  As a
matter of principle, however, any nonlinear recurrence can be seen as
an efficient implementation of a large order linear recurrence. The
popular ``Mersenne Twister''
\cite{Matsumoto:Nishimura:1998:Mersenne_Twister} for example is an
efficient nonlinear implementation of a linear recurrence in $\F_2$ of order
19\,937.

The coefficients and the initial seed of the minimum order linear
recurrence can be calculated very efficiently, i.e., polynomially in the
minimum order, using the Berlekamp-Massey algorithm
\cite{Jungnickel:1993:Finite_Fields}.

\begin{figure}
  \centering
  \includegraphics[width=0.7\linewidth]{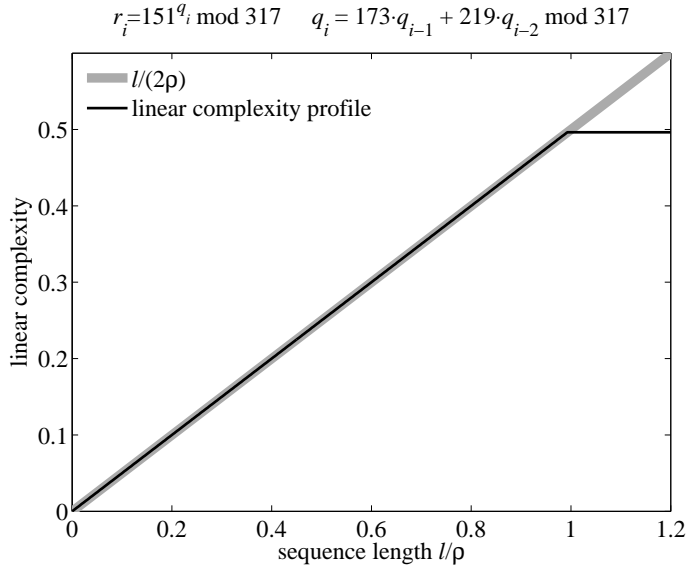}
  \caption{Linear complexity profile $L_{(r)}(l)$ of a YARN sequence, produced
    by the recurrence $q_i = 173\cdot q_{i-1} + 219 \cdot q_{i-2}\bmod 317$
    and $r_i = 151^{q_i} \bmod 317$. The period of this sequence is
    $T=317^2-1$.
  \label{fig:linear_complexity_profile}}
\end{figure}

If we feed a stream of numbers from a PRNG to the Berlekamp-Massey
algorithm, we get its linear complexity as a function of the length
$\ell$ of the stream.  For a LFSR sequence \eqref{eq:lfsr}, this
linear complexity profile quickly reaches a plateau at $\ell=n$,
whereas for a truly random stream we expect the profile to grow like
$\ell/2$.  Figure~\ref{fig:linear_complexity_profile} shows that the
linear complexity of a YARN generator does grow like $\ell/2$ for
$\ell \leq T/2$. Measured in terms of linear complexity, YARN generators
are as random and nonlinear as it can get.

\section{Parallelization}
\label{sec:parallelization}

In parallel applications we need to generate streams $t_{j,i}$ of random
numbers, where $j=0,1,\dots,p-1$ numbers the streams for each of the $p$
processes. We require statistical independency of the $t_{j,i}$ within each
stream and between the streams. 

Since random numbers are very often needed in the innermost loop of a
simulation, efficiency of a PRNG is very important.  We require that a
PRNG can be parallelized without loosing its efficiency.

And there is a third requirement besides quality and speed: we want the
simulation to \emph{play fair}. We say that a parallel Monte Carlo simulation
plays fair, if its outcome is strictly independent of the underlying
hardware, where strictly means deterministically, not just
statistically. Fair play implies the use of the same pseudo random
numbers in the same context, independently of the number of parallel
processes. It is mandatory for debugging, especially in parallel
environments where the number of parallel processes varies from run to
run, but another benefit of playing fair is even more important: Fair
play guarantees that the quality of a PRNG with respect to an
application does not depend on the degree of parallelization. In this sense,
fair play conserves quality.

It is not always easy to implement a Monte Carlo simulation such that
it plays fair. For some Monte Carlo algorithms it may even be impossible.
But for many simulation algorithms it is possible---provided the
underlying PRNG supports fair parallelization. We will see that some
straightforward parallelization techniques for PRNGs prevent fair play
even in the simplest simulation, while others support fair play
in moderately complicated situations.

\subsection{Random Seeding and Parameterization}

One parallelization technique is \emph{random seeding}: All processes
use the same PRNG but a different ``random'' seed. The hope is that
they will generate non-overlapping and uncorrelated subsequences of
the original PRNG.  This hope, however, has little theoretical
foundation.  Random seeding is a clear violation of Don Knuth's advice
cited above. Yet this approach is frequently used in practice because it works
with any PRNG, is easy to program and comes without any penalty on the
efficiency. With random seeding, however, no simulation can play fair.

Another parallelization technique is \emph{parameterization}: All
processes use the same type of generator but with different parameters
for each processor. Example: linear congruential generators with
additive constant $b_j$ for the $j$th stream
\begin{equation}
  t_{j,i}=a\cdot t_{j, i-1} + b_j \bmod m \,,
\end{equation}
where the $b_j$'s are different prime numbers just below $\sqrt{m/2}$.
Another variant uses different multipliers $a$ for different streams.
The theoretical foundation of these methods is weak, and empirical
tests have revealed serious correlations between streams. On massive
parallel system you may need thousands of parallel streams, and it is
not trivial to find a type of PRNG with thousands of ``well tested''
parameter sets.

Like random seeding, parameterization prevents fair play. Therefore
both methods should be avoided.

\subsection{Block Splitting and Leapfrogging}

\emph{Block splitting} is the idea to break up a single stream of
random numbers into non-overlapping, consecutive blocks of numbers and
assign each block to one of the processes.

\begin{figure}
  \centering
    \includegraphics[width=0.46\linewidth]{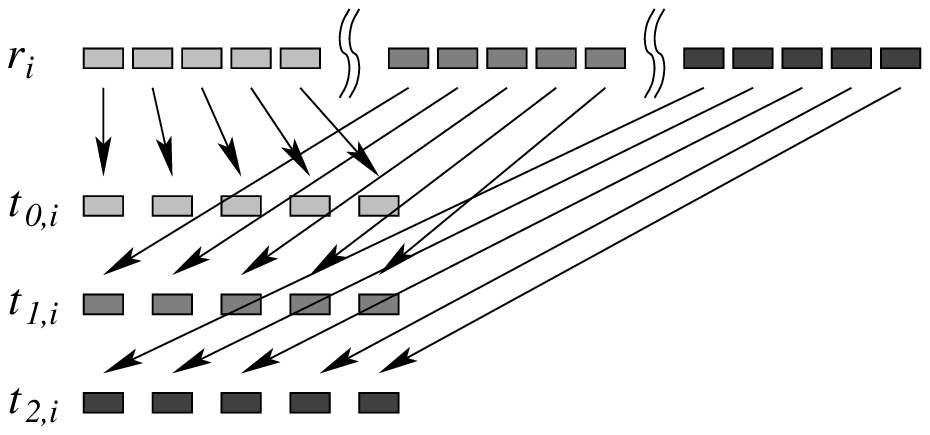}
    \includegraphics[width=0.46\linewidth]{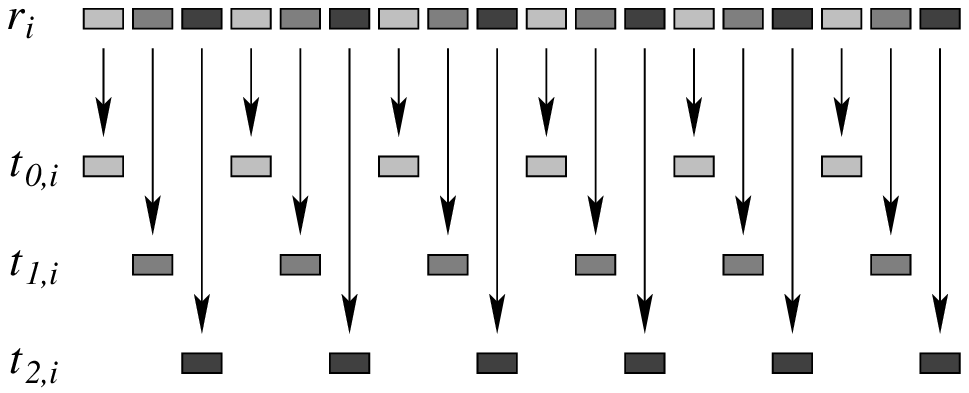}
    \caption{Parallelization by block splitting (left) and leapfrogging (right).
    \label{fig:blocks}}
\end{figure}

Let $M$ be the maximum number of calls to a PRNG by each processor,
and let $p$ be the number of processes.  Then we can split the
sequence $(r)$ of a sequential PRNG into consecutive blocks of length
$M$ such that
\begin{equation}
  \begin{split}
    t_{0, i} & = r_{i} \\
    t_{1, i} & = r_{i+M} \\
    & \dots\\
    t_{p-1, i} & = r_{i+M(p-1)}\,.
  \end{split}
\end{equation}
This method works only if we know $M$ in advance or can at least
safely estimate an upper bound for $M$. For an efficient block splitting, it is
necessary to jump from the $i$th random number to the $(i+M)$th number
without calculating the numbers in between.
Block splitting is illustrated in Figure~\ref{fig:blocks}.

The \emph{leapfrog} method distributes a sequence $(r)$ of random
numbers over $d$ processes by decimating this base sequence such that
\begin{equation}
  \label{eq:leap_frog}
  \begin{split}
    t_{0, i} & = r_{di} \\
    t_{1, i} & = r_{di+1} \\
    & \dots\\
    t_{d-1, i} & = r_{di+(d-1)}\,.
  \end{split}
\end{equation}
Leapfrogging is illustrated in \figurename~\ref{fig:blocks}. It does
not require an a priori estimate of how many random numbers will be
consumed by each processor.

Note that for a periodic sequence $(r)$ the substreams derived from
block-splitting are cyclic shifts of the original sequence $(r)$, and
for $p$ not dividing the period of $(r)$, the leapfrog sequences are
cyclic shifts of each other. Hence the leapfrog method is equivalent
to block-splitting on a different base sequence.

Leapfrog and block splitting support fair play, i.e., they allow us to 
program parallel simulations that use the same random numbers in the same
context idependently of the number of parallel streams. 

As an illustrative example we consider the site percolation problem. A
site in a lattice of size $N$ is occupied with some probability, and
the occupancy is determined by a pseudo random number. $M$ random
configurations are generated.  

A fair playing percolation simulation can be organized by
leapfrogging on the level of lattice configurations. Here each process
consumes distinct contiguous blocks of numbers from the sequence $(r)$, and the
workload is spread over $d$ processors in such a way that each process
analyses each $d$th lattice. If we number the processes by their rank $i$ from
$0$ to $d-1$ and the lattices from $0$ to $M-1$, each process starts with a
lattice whose number equals its own rank. That means process $i$ has to skip
$i\cdot N$ random numbers before the first lattice configuration is generated.
Thereafter each process can skip $d-1$ lattices, i.\,e., $(d-1)\cdot N$
random numbers, and continue with the next lattice.

Organizing simulation algorithms such that they play fair is not
always as easy as in the above example, but with a little effort one
can achieve fair play in more complicated situations, too. This may
require the combination of block splitting and the leapfrog method, or
iterated leapfrogging.  Sometimes it is also necessary to use more
than one stream of random numbers per process, as we will see below.

\subsection{Parallelization of Linear Recurrences}

We can ``simulate'' block splitting and leapfrogging by throwing away
all random numbers that are not from the right block or from the right
leapfrog subsequence. This works for any PRNG, but it is not very
efficient and foils parallelization. For an efficient parallelization,
we should be able to modify a PRNG to generate directly only every
$d$th element of the original sequence (for leapfrogging) or to
directly advance the PRNG by $M$ steps (for block splitting).
For LFSR sequences \eqref{eq:lfsr}, both can be achieved very efficiently.

Let's start with block splitting, i.e., with jumping ahead in a linear
recurrence. Note that by introducing a companion matrix $A$ the
linear recurrence \eqref{eq:lfsr} can be written as a vector matrix
product:
\begin{equation}
  \begin{pmatrix}
    r_{i-(n-1)} \\ \vdots \\ r_{i-1} \\ r_{i}
  \end{pmatrix}=\underbrace{
  \begin{pmatrix}
    0 & 1 & \dots & 0 \\
    \vdots & \vdots & \ddots & \vdots \\
    0 & 0 & \dots & 1 \\
    a_n & a_{n-1} & \dots & a_1 
  \end{pmatrix}}_{\displaystyle A}
  \begin{pmatrix}
    r_{i-n} \\ \vdots \\ r_{i-2} \\ r_{i-1}
  \end{pmatrix}
  \bmod p
\end{equation}
From this formula it follows immediately that the $M$-fold successive
iteration of \eqref{eq:lfsr} may be written as
\begin{equation}
  \begin{pmatrix}
    r_{i-(n-1)} \\ \vdots \\ r_{i-1} \\ r_{i}
  \end{pmatrix}=
  A^M
  \begin{pmatrix}
    r_{i-M-(n-1)} \\ \vdots \\ r_{i-M-1} \\ r_{i-M}
  \end{pmatrix}
  \bmod p\,.
\end{equation}
Matrix exponentiation can be accomplished in $\bigO{n^3\ln M}$ steps via
binary exponentiation, also known as exponentiation by squaring.

Implementing leapfrogging efficiently is less straightforward.
Calculating $t_{j,i}=r_{di+j}$ via
\begin{equation}
  \begin{pmatrix}
    r_{di+j-(n-1)} \\ \vdots \\ r_{di+j-1} \\ r_{di+j}
  \end{pmatrix}=
  A^d
  \begin{pmatrix}
    r_{d(i-1)+j-(n-1)} \\ \vdots \\ r_{d(i-1)+j-1} \\ r_{d(i-1)+j}
  \end{pmatrix}
  \bmod m
\end{equation}
is no option, because $A^d$ is usually a dense matrix, in which case
calculating a new element from the leapfrog sequence requires $\bigo(n^2)$
operations instead of $\bigo(n)$ operations as in the base sequence.

Obviously the linear complexity of a leapfrog
subsequence cannot be larger than the linear complexity of the
underlying base sequence, which for LFSR sequences is $n$. Hence all
we need to do is to generate at most $2n$ elements of the leapfrog
sequence and feed them to the Berlekamp-Massey algorithm. This yields
the coefficients and the seed for an LFSR generator that produces the
leapfrog subsequence. This is how leapfrogging for LFSR sequences is done
in practice.

Note that the techniques for block splitting and for leapfrogging of
LFSR sequences also work for YARN generators by simply applying them
to the underling LFSR sequence.

\section{From Theory to Practice: TRNG}

We will illustrate the use of pseudo random generators in parallel 
simulations by means of TRNG, a C++ library for sequential and
parallel Monte-Carlo simulations. TRNG has several outstanding features 
that set it apart from other random number libraries:
\begin{enumerate}
\item Its current version (4.7) contains a collection of different
  PRNGs, like various LFSR and YARN generators, lagged Fibonacci
  sequences, 32-bit and 64-bit implementations of the Mersenne-Twister,
  etc.
\item Its C++ interface makes it very easy to switch PRNGs in your
  simulation, even if your program is completely written in C (see
  below).
\item Parallelization by leapfrogging and block-splitting is fully
  supported for all generators that can be parallelized.
\item The internal state of each PRNG can be written to and read from a
  file. This allows us to stop a simulation and carry on later, which
  is especially useful for long running simulations that face the risk
  of hardware failure or system maintenance before they are done.
\item TRNG implements a large variety of distributions, from standard
  distributions like uniform, exponential or Poisson distributions to
  lesser known distributions like Fisher-Snedecor- and
  Rayleigh-distributions. Each distribution comes with functions
  to calculate its probability density, its
  cumulative distribution and, in the case of continuous
  distributions, its inverse cumulative distribution.
\item The interface of TRNG complies to the ISO C++ standard for PRNGs
  \cite{Brown:Fischler:Kowalkowski:Paterno:2006:Comprehensive_Proposal}
  \emph{and} to the random number generator interface defined in the
  Standard Template Library (STL) for C++ \cite{josuttis:stl}.
\item Last but not least: TRNG is free. You can get the source code and the documentation from \url{trng.berlios.de}.
\end{enumerate}

\subsection{Basics}

TRNG is a set of classes of two types: random number engines
(the actual PRNGs) and probability distributions.
Each class is defined in its own header file.
Lets suppose that we want to use random number generators of
the YARN type (for the parallel part) and a Mersenne twister generator for the sequential part. We will generate uniform variates from the interval $[0,1)$ and Poisson variates. The corresponding header files read
{
\small
\begin{verbatim}
   #include <trng/yarn2.hpp>
   #include <trng/mt19937.hpp>
   #include <trng/uniform01_dist.hpp>
   #include <trng/poisson_dist.hpp>
\end{verbatim}
}
\noindent
and the PRNGs and distributions are declared as
{
\small
\begin{verbatim}
   trng::yarn2 rand1, rand2;  // two generators of type YARN
   trng::mt19937 rand3;       // Mersenne twister
   trng::uniform01_dist<double> uniform;
   trng::poisson_dist poisson(2.0);  // mean value is 2.0
\end{verbatim}
}
Note that we specified the uniform distribution to return numbers of type
\texttt{double}. Alternatively we could have specified \texttt{float} or
\texttt{long double}. The Poisson distribution returns integer values, of course. 

The '2' in the classname \texttt{yarn2} denotes the order of the
underlying LFSR. The feedback parameters are
set to well chosen default values, but TRNG allows users to supply
their own paremeters.

The use of the generators is very simple:
{
\small
\begin{verbatim}
   k = poisson(rand3);  // use Mersenne twister to generate Poisson variate
   ...
   if (uniform(rand1) < 0.5) {
     // do this with probability one half
     ...
   }
\end{verbatim}
}

Most of the random number engines in TRNG are parallelizable, i.e.,
they have member functions for leapfrogging and
block-splitting. Let \texttt{nprocs} denote the
total number of parallel processes, and let \texttt{rank} denote the
ordinal number of the current process. A parallel simulation code
typically contains lines like
{
\small
\begin{verbatim}
  rand1.split(nprocs,rank);
\end{verbatim}
} This line modifies the random number engine \texttt{rand1} such that
it generates leapfrog substream \texttt{rank} out of
\texttt{nprocs} total substreams.

For block-splitting there exist two functions:
{
\small
\begin{verbatim}
  rand1.jump(M); // advance by M steps
  rand2.jump1(n); // advance by 2^n steps
\end{verbatim}
}

\subsection{Example: Broken Triangles}

To see how TRNG is used in a concrete situation, let's 
develop a parallel Monte Carlo code to analyse the following problem:
\begin{quote}
  If two points are independently and uniformly located in the unit
  interval, they divide that interval into three segments. What is the
  probability that those three segments form a triangle? And what is
  the probability that the triangle is obtuse?
\end{quote}
The mathematical background of this problem is very simple: Line
segments of lengths $a$, $b$ and $c$ can be arranged as a triangle if
and only if all lengths obey the triangle inequality, i.e., if and
only if
\begin{equation}
  \label{eq:triangle-inequality}
  a \leq b + c \qquad b \leq a+c \qquad c \leq a+b\,.
\end{equation}
We recall from high school geometry that the interior angles of a
triangle $\alpha$, $\beta$ and $\gamma$ are given by
\begin{equation}
  \label{eq:triangle-angles}
  \cos\alpha = \frac{b^2+c^2-a^2}{2bc}, \qquad
  \cos\beta = \frac{a^2+c^2-b^2}{2ac}, \qquad
  \cos\gamma = \frac{a^2+b^2-c^2}{2ab}\,.
\end{equation}
A triangle is obtuse if one of the angles is larger than $90°$, i.e.,
if one of the nominators is negative.

In our Monte Carlo algorithm for the triangle problem, we
generate two random numbers $r_1$ and $r_2$ uniformly
and independently from the unit interval and calculate the lengths of
the segments
\begin{displaymath}
  a = \min(r_1,r_2) \qquad b = \max(r_1,r_2) - \min(r_1,r_2) \qquad 
  c = 1 - \max(r_1,r_2)\,.
\end{displaymath}
Then we check \eqref{eq:triangle-inequality} and the nominators of
\eqref{eq:triangle-angles} to see whether the segments form an
(obtuse) triangle. We repeat this experiment over and over again and count the
fraction of (obtuse) triangles. 

Once again we assume that the number of running processes and
the rank of a process are stored in the variables \texttt{nprocs} 
and \texttt{rank}.
The core part of our parallel simulation then reads 
{
  \small
\begin{verbatim}
  trng::yarn2 r1, r2;

  r1.seed (seed); 

  // first splitting yields two random sources for the two cuts
  r2 = r1;
  r1.split (2,0); r2.split (2,1); // ... 

  // second splitting for parallelization
  r1.split (nprocs,rank); r2.split (nprocs,rank);

  // main loop
  triangles = 0; obtuse = 0;
  for (k = 0; k < MCS; k += nprocs) { // MCS: number of total samples
    a = uniform(r1);
    b = uniform(r2);
    if (a > b) {c = a; a = b; b = c;} // ensure a <= b
    c = 1.0 - b;
    b = b-a;
    if ((a+b >= c) && (a+c >= b) && (b+c >= a)) {
      triangles++;
      if ((a*a+b*b < c*c) || (a*a+c*c < b*b) || (b*b+c*c < a*a)) obtuse++;
    }
  }
\end{verbatim}
}
We used leapfrogging twice. The first split creates two sequences,
one for each random cut into the unit interval. Then both sequences are
split according to the number of parallel processes. The loop is written
such that each process gets the same workload.

The complete program has additional code for getting the values for
\texttt{nprocs} and \texttt{rank}, for distributing the seed 
among all processes and for collecting the results from all processes.
Since this part depends on the parallel machinery (like MPI or
OpenMP), we will not discuss it here. But you are encouraged to
complete the program and run it.

Here is the output of a MPI implementation running on 20 processes,
with seed $141164$ and using $10000$ samples:
{
  \small
\begin{verbatim}
  /home/mertens> mpirun -np 20 mpi_triangle -S 141164 -M 10000
  # number of processes: 20
  # PRNG: yarn2 from TRNG 4.7
  # seed    = 141164
  # samples = 10000
  # fraction of triangles: 0.2481
  # fraction of obtuse triangles: 0.1725
\end{verbatim}
}
\noindent
This result is within the error bars of the exact values
\begin{displaymath}
  p(\text{triangle}) = \frac{1}{4} \qquad p(\text{obtuse triangle}) = \frac{9}{4}-3\ln 2 = 0.170558\ldots\,,
\end{displaymath}
hence we are confident that our program is correct. But if we run it with exactly the same parameters on 30 processes, we get
{
  \small
\begin{verbatim}
  /home/mertens> mpirun -np 30 mpi_triangle -S 141164 -M 10000
  # number of processes: 30
  # PRNG: yarn2 from TRNG 4.7
  # seed    = 141164
  # samples = 10000
  # fraction of triangles: 0.2486
  # fraction of obtuse triangles: 0.1728
\end{verbatim}
}
\noindent
The deviation from the previous run is small and well within the
statistical error. But wait! We are running a fair simulation. The
output must not depend on the number of parallel processes. The tiny
deviation tells us that our program is not correct. Try to find the
bug in the code above!

\subsection{Final Remarks}

``Which random number generator is the best? Which one shall I use in
my simulations?'' These questions are inevitably asked after each and
every talk on PRNGs. The answer is simple: There is no such thing as
the ``best'' generator. Most generators are advertised with a phrase
like ``this generator has passed a battery of statistical tests''.
But a generator that has passed $N-1$ statistical tests may fail the
$N$th test---your simulation. And LFSR sequences that
fail the spectral test, work perfectly in most applications that do
not sample points in high dimensional space. The rule is, that your
simulation should not be sensitive to the inevitable correlations in a
PRNG  \cite{bauke:mertens:rng2}. But how can we know? By following
rule number one:
\begin{description}
\item[Rule I:] Run your simulation at least twice, each time with a different
type of PRNG.
\end{description}
If the results of all runs agree within the statistical error, we are on the safe side.

In order to enable others to repeat your numerical experiments, the type of
PRNG must be published along with the results. In addition, the PRNG and the
seed used in the run should be stored with the data:
\begin{description}
\item[Rule II:] The type of PRNG that has been used must be published
  along with the results. The seed must be stored in the original data files.
\end{description}
Quite often, Monte Carlo simulations screw up not because of a bad
PRNG but because of the bad usage of a PRNG. The classical example is
to (mis)use a $32$-bit random integer to sample a list of $2^{36}$
elements. Such a list corresponds to $64$ GBytes, which is not an
uncommon size. This brings us to
\begin{description}
\item[Rule III:] Know your PRNG, especially its limitations (numerical
resolution, period length, statistical quality of low order bits, etc.).
\end{description}
Last but not least:
\begin{description}
\item[Rule IV:] Parallel simulations should be programmed to play fair, and fair play should explicitely be checked.
\end{description}

\section{Further Reading}

This lecture is mainly based on the paper that introduces the YARN generator and the concept
of fair play in Monte-Carlo simulations
\cite{bauke:mertens:rng3}. The references in this paper can be used as a starting point
for a journey into the wondrous world of pseudo randomness. Another good starting point
is Brian Hayes' nice essay ``Randomness as a Resource'' \cite{Hayes:2001:Randomness}.

If you want to learn more about the theory behind pseudo random number
generators, Donald Knuth's classical tome \cite{Knuth:1998:TAoCPII} is
the right choice.  An excellent introduction to finite fields is
Jungnickel \cite{Jungnickel:1993:Finite_Fields}.

Every physicist who is using random number generators should know about the
\emph{Ferrenberg affair}: In 1992, Alan Ferrenberg, David Landau and Joanna Wong discovered
that a family of established random number generators yield wrong results in
Monte Carlo simulations of the Ising model \cite{Ferrenberg:Landau:1992}. This
was a shock to the community because up to this moment, everybody trusted PRNGs that
had passed the infamous ``battery of statistical tests''. Read Brian Hayes' depiction
of the Ferrenberg affair and its aftermath \cite{Hayes:1993:Wheel}. The forensic
analysis of the affair is also recommended reading \cite{mertens:bauke:rng1}.

If you want to see how (and why) well established PRNGs fail dramatically even in very
simple simulations, you will enjoy \cite{bauke:mertens:rng2}.

\end{document}